\documentclass[]{raa} 

\newcommand*{\dif}{\mathop{}\!\mathrm{d}}

\newcommand{\sref}[1]{Sect.~\ref{#1}}

\newcommand{\fref}[1]{Fig.~\ref{#1}}

\newcommand{\eref}[1]{Eq.~\eqref{#1}}

\newcommand{\tref}[1]{Table~\ref{#1}}

\usepackage{graphicx,times}
\usepackage{natbib}
\usepackage{hyperref}
\usepackage{amssymb}
\usepackage{amsmath}
\bibpunct{(}{)}{;}{a}{}{,}

\usepackage{tikz}
\usetikzlibrary{shapes.geometric, arrows}
\usetikzlibrary{positioning}

\tikzstyle{startstop} = [rectangle, rounded corners, minimum width=3cm, minimum height=1cm,text centered, draw=black, fill=red!30]
\tikzstyle{process} = [rectangle, minimum width=3cm, minimum height=1cm, text centered, draw=black, fill=orange!30]
\tikzstyle{arrow} = [thick,->,>=stealth]

\begin{document}

\title{Measuring Tianlai’s primary beam using sky model}

   \volnopage{Vol.0 (20xx) No.0, 000--000}      
   \setcounter{page}{1}          

   \author{Yunbo Geng 
      \inst{1,2}
   \and Furen Deng
      \inst{1,2}
   \and Jixia Li
      \inst{1,2}
    \and Shifan Zuo
        \inst{1}
    \and Shijie Sun
        \inst{1}
    \and Yichao Li
        \inst{3}
    \and Fengquan Wu
         \inst{1} 
    \and Yougang Wang
        \inst{1,2,3}      
    \and Xuelei Chen
        \inst{1,2,3}
   }

   \institute{
    State Key Laboratory of Radio Astronomy and Technology, National Astronomical Observatories, CAS, A20 Datun Road, Chaoyang District, Beijing, 100101, P. R. China; {\it wangyg@bao.ac.cn, \it xuelei@bao.ac.cn}\\
        \and
    School of Astronomy and Space Science, University of Chinese Academy of Sciences, Huairou District, Beijing 101408, P. R. China\\
        \and
    Key Laboratory of Cosmology and Astrophysics (Liaoning) \& College of Sciences, Northeastern University, Shenyang 110819, China\\
\vs\no
   {\small Received 20xx month day; accepted 20xx month day}}

\abstract{We present the beam pattern measurement of the Tianlai Cylinder Pathfinder Array. As it is a pure drift-scan instrument, we exploit the North-South motion of the Sun to demonstrate that the primary beam is factorizable. Leveraging this property, we decompose the primary beam into independent East-West (E-W) and North-South (N-S) components. Using the Sun as a calibration source, we obtain the E-W beam profiles at various elevations, applying normalization to eliminate the effects of solar activity. Subsequently, we simulate the observed signals using a sky map model to derive the best-fit N-S beam. The results of this work are consistent with previous expectations.}

\authorrunning{Y. Geng, F. Deng, J. Li, Y. Wang \& X. Chen }            
\titlerunning{Measuring Tianlai’s primary beam using sky model}  
\maketitle



\section{Introduction}
\label{section:introduction}
The Tianlai (Chinese for “heavenly sound”) project \citep{Chenxl2012tianlai,Xu2014-jr} is a radio interferometric array.
It aims to observe the redshifted 21 cm line of neutral hydrogen and utilize the intensity mapping technique to obtain the 3D distribution of matter \citep{PhysRevLett.100.091303,Ansari2008}. The 21 cm line emitted by neutral hydrogen serves as a tracer of matter distribution, enabling us to probe the expansion history of the universe and the properties of dark matter and dark energy. 
Currently, the project is in the Pathfinder stage,  which comprises two distinct arrays: a dish array of 16 dish radio telescopes  \citep{wu/10.1093/mnras/stab1802, Zhang2016-im} and a cylinder array. The latter consists of three cylinder reflectors with a total of 96 dual polarization receivers \citep{LiCylDes, Zhang2016-cc}. 

A critical challenge in analyzing data from the Tianlai Cylinder Pathfinder Array (TCPA) is the uncertainty in its primary beam pattern.
Although electromagnetic simulations can provide a theoretical beam model \citep{Sun_2022},  the antenna's structural complexity can lead to significant inaccuracies. Unlike dish antennas with symmetric feeds, which often have azimuthally symmetric beams, the beam of a fixed cylindrical transit telescope is more complex. Its East-West (E-W) response is governed by the array factor and electronic phasing, varying with frequency and pointing, whereas the North-South (N-S) response is determined by the fixed mechanical structure. This complexity makes empirical calibration challenging.

Traditional calibration methods have limitations. One can observe transiting astrophysical point sources to derive the E-W beam cut, but this is only possible at the few declinations (zenith angles) of sufficiently bright sources. This issue is common among all fixed-transit telescopes, particularly cylinder arrays like the  Molonglo Observatory Synthesis Telescope (MOST) \citep{most}, the Canadian Hydrogen Intensity Mapping Experiment (CHIME) \citep{CHIME}, and the TCPA itself. 
Another approach involves using artificial sources carried by unmanned aerial vehicles (UAVs or drones) \citep{Kuhn2025-tl}.  This method has been successfully applied to the Tianlai Dish array \citep{9638547} and TCPA  \citep{LiDrone}. However, accurately satisfying the far-field condition with drones is difficult, and discrepancies remain between drone-based measurements and simulations \citep{Sun_2022}.

Recently, the CHIME team \citep{chimebeam} demonstrated a novel technique using the Sun as a calibration source for primary beam measurement. By beamforming on six months of data and calibrating with daily solar flux, they derived the array's response across the range of solar declinations. As the Sun moves between its northern and southern declination limits over the year, this method provides high signal-to-noise ratio (SNR) measurements of the beam across a wide range of N-S angles. However, this approach is not directly transferable to the TCPA.
Its success relies on the presence of redundant baselines for internal consistency and beam mapping, whereas the feeds on the Tianlai cylinders are deliberately arranged in a non-redundant configuration to minimize grating lobes \citep{Zhang_2016}. Additionally, accurately correcting for the intrinsic variability of solar radio flux remains a significant challenge.

In this work, we introduce a novel and independent method for measuring the TCPA's primary beam. Our approach is based on the assumption that the beam is separable in the N-S and E-W directions. We then measure these two components independently. First, we use the Sun as a calibrator to measure the E-W beam profile. Second, by simulating the array's output signals using existing sky survey models, we employ a Markov Chain Monte Carlo (MCMC) analysis to constrain the most probable N-S beam profile. Finally, we validate the initial separability assumption by comparing our reconstructed beam with previous measurements and simulations.

Below, we introduce our method used for primary beam measurement in \sref{section:methods}. We present our findings in \sref{section:results}, where we also design tests to verify our assumptions. The error analysis and limitations of our method are discussed in \sref{section:discussion}. Finally, we summarize our results in \sref{section:conclusions}.

\section{Methods}
\label{section:methods}

\subsection{The Tianlai Cylinder}


\begin{figure}[htbp]
    \centering
    \includegraphics[width=0.6\textwidth]{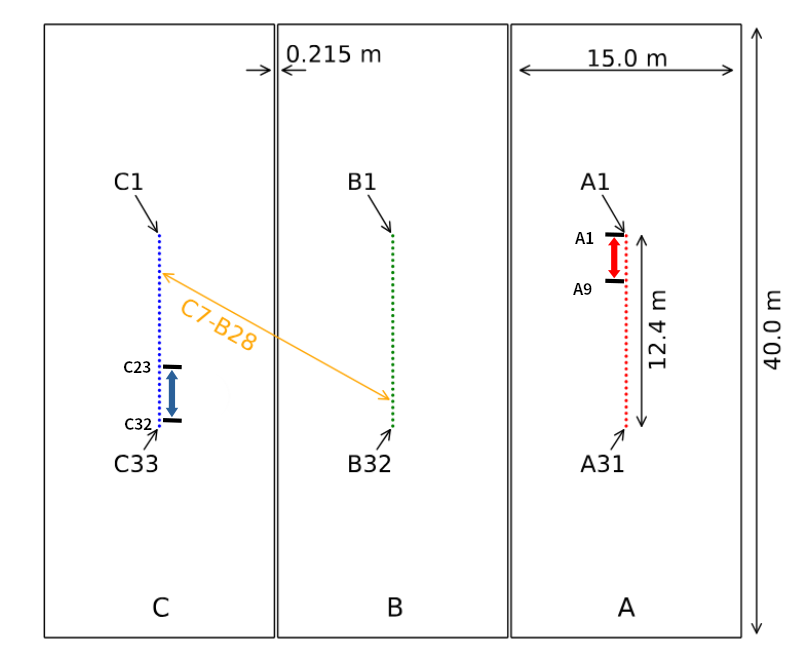} 
    \caption{Schematic layout of the Tianlai Cylinder Pathfinder Array (TCPA). The array consists of three cylinders (from right to left) containing 31, 32, and 33 feeds, respectively. Their positions are illustrated in this diagram. Baseline A1–A9 and Baseline C23–C32 are shown as representative examples for the North-South and East-West primary beam measurements, respectively. For clarity, other baselines used in the work are omitted to prevent the overlapping of markers. \citep{LiCylDes}
    }
    \label{fig:tianlaicyl}
\end{figure}

The structural configuration of the TCPA is illustrated in Fig.\ref{fig:tianlaicyl}. Each cylinder measures 40 m in length and 15 m in width, with dual-polarization dipole feeds installed at the focal line \citep{tianlaifeed}. The TCPA currently operates within a frequency range of 700–800 MHz. It features a wide field of view (FoV) in the North-South direction, while its East-West FoV is relatively narrow.

\begin{figure}[htbp]
    \centering
    \begin{tikzpicture}[node distance=0.5cm]

        \node (start) [startstop] {Input Data};
        \node (pro1) [process, right=of start] {Detect Noise Source};
        \node (pro11) [process, right=of pro1] {RFI Flagging};
        \node (pro2) [process, right=of pro11] {Strong Source Calibration};
        \node (pro3) [process, below=of start, yshift=-0.5cm] {Relative Phase Calibration};
        \node (pro4) [process, right=of pro3] {Detect Bad Baseline};
        \node (pro5) [process, right=of pro4] {Subtract the nightly mean};
        \node (stop) [startstop, right=of pro5] {Output Data};

        \draw [arrow] (start) -- (pro1);
        \draw [arrow] (pro1) -- (pro11);
        \draw [arrow] (pro11) -- (pro2);
        \draw [arrow] (pro2) -- ++(2,0) |- ++(-15,-1) |- (pro3);
        \draw [arrow] (pro3) -- (pro4);
        \draw [arrow] (pro4) -- (pro5);
        \draw [arrow] (pro5) -- (stop);
    \end{tikzpicture} 
    \caption{Data reduction pipeline for the raw observations. }
    \label{fig:tianlai_outline}
\end{figure}
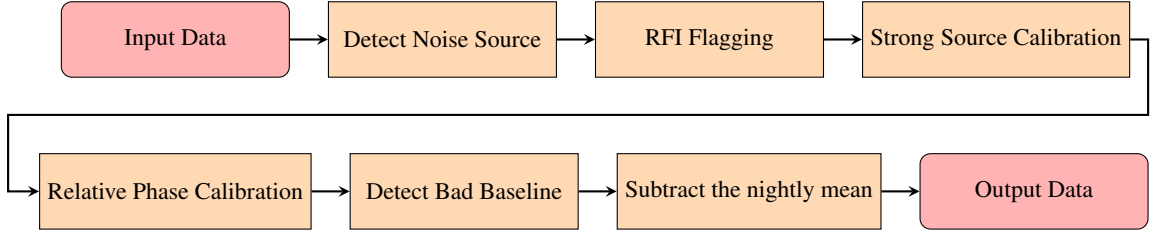

The system generates visibility data over time, which represent the correlation product of the voltages between each pair of feeds. Signals from the feeds are amplified and transmitted to the back-end via optical fibers. At the back-end, they are further amplified by analog circuits processed through the F-engine to separate frequencies via Fast Fourier Transform (FFT), and finally correlated by the X-engine to produce the output. 
The raw data are subsequently processed through a pipeline—including steps such as 
 Detect Noise Source used for calibration, 
Radio Frequency Interference (RFI) flagging, calibration, and subtraction of the night-time mean—using the \texttt{tlpipe} software package, which is described in detail in \citet{tlpipe}. The complete data processing workflow is depicted in \fref{fig:tianlai_outline}. Notably, this methodology can leverage existing data sets from other Tianlai observations, eliminating the need for dedicated separate observations. The specific data sets utilized in this study are summarized in \tref{tab:testdata}.

\subsection{Primary Beam Model}

We model the primary beam of the TCPA as a product of a function along the North-South direction and a function along the East-West, which is natural for cylinders \citep{Shaw2015,Yusimu},
\begin{equation}
    \label{eq2:beamodel}
    A(\vec{n}) = A_\mathrm{NS}(\sin^{-1}(\vec{n}\cdot\vec{x})) \times A_\mathrm{EW}(\sin^{-1}(\vec{n}\cdot\vec{y}))
\end{equation}
where $\vec{x}$ and $\vec{y}$ are the unit vectors pointing East and North. We assume the beam factor along the North-South direction has a Gaussian form \citep{Yusimu}
\begin{equation}
    \label{eq2:beamNS}
    A_\mathrm{NS}(\theta) = \exp\left(- \frac{\theta^2}{\theta_\mathrm{NS}^{2}}\right)
\end{equation}
with a parameter $\theta_\mathrm{NS}$, and the beam factor along  the East-West direction has the form \citep{Yusimu}
\begin{equation}
    \label{eq2:beamEW}
    A_{\rm EW}(\theta) = \int_{-\frac{W}{2}}^{\frac{W}{2}}A_{D}\left(2\tan^{-1}(\frac{x}{2FW}), \theta_{\rm EW}\right) e^{-ikx\sin\theta} \dif x
\end{equation}
where $F$ is the focal ratio, $W$ is the width of the cylinder, and $A_D$ is taken as\citep{Yusimu}
\begin{equation}
    \label{eq2:beamD}
    A_{D}(\theta; \theta_{\rm EW}) = e^{- \frac{\ln2}{2}\frac{\tan^2\theta}{\tan^2(\frac{\theta_{\rm EW}}{2})}}
\end{equation}
This model approximates the results obtained from the electromagnetic (EM) field simulation by \cite{Sun_2022}, as thoroughly validated in the work of \cite{Yusimu}.

\subsection{East-West Beam Measurement}
The sensitivity limits of the Tianlai array result in a sparse population of radio sources suitable for primary beam measurement in the East-West (E-W ) direction. 
Therefore, we chose the Sun as the source for measuring the E-W beam. The Sun's motion between the Tropics of Cancer and Capricorn provides us with a larger number of data points. This also allows us to validate our model's assumption that the primary beam is separable. Furthermore, as the Sun is a strong radio source, its signal is bright enough to be largely unaffected by confusion from other background sources.

Given the inherently narrow primary beam of the TCPA in the E-W direction, we can reasonably assume that the angle in the North-South direction of the Sun within the primary beam, represented by $\sin^{-1}(\vec{n}\cdot\vec{x})$, remains constant. We adopt the method described in \cite{chimebeam} to measure the primary beam in the E-W direction. Finally, by utilizing solar transit data collected on different days, we validated the assumption of separability in our primary beam model.

\subsubsection{Baseline Selection}

 The structure diagram of the TCPA is shown in \fref{fig:tianlaicyl}. It consists of three cylindrical parabolic reflectors, with the feeds arranged along the focus line of the cylindrical reflectors. Baseline is a vector from one feed to another, such as baseline C7-B28 in \fref{fig:tianlaicyl}, which is a vector between feed C7 and feed B28.

To calibrate the beam profile along the North-South direction, we need a bright point source which moves in declination. The Sun moves along the ecliptic circle, thus its declination changes over the season. However, the Sun is not a point source. 
The quiet Sun in the radio band is well described by a disk approximately $0.5^{\circ}$ in diameter \citep{sunimage}. Sometimes the Sun has active regions which generate much stronger radio emission, which varies with intensity and spectrum, and the shape, size and  position of such region also varies. Nonetheless,  if the baseline is short, then the corresponding angular resolution would not be sufficient to resolve the Sun or its emitting regions, and it can be regarded effectively as a point source.  Furthermore, while solar radio bursts may induce fluctuations in received signal strength, for most solar bursts most energy is emitted at lower frequencies\citep{White2024-jo}, while for our observation band at 700-800 MHz the steady emission from the quiet Sun is more common \citep{Vasanth2025-dw}. 

The angular resolution of the baseline is $\theta = \frac{\lambda}{L}$, where $L$ is the baseline length. Here, we select baselines with $\theta > 1^{\circ}$, so the Sun can be treated as a point source. The center frequency of the TCPA is 750 MHz, corresponding to a wavelength $\lambda$ of 0.4 m, the baseline length $L$ should ideally not exceed 22.9 m to satisfy this requirement. 
The properties of each cylinder are slightly different. In the present work, we selected the baselines formed by feeds ranging from C13 to C33 for illustration. To prevent cross-coupling between feeds, we selected those with $L > 1.8\text{m}$. At the same time, the constraint $L < 4\text{m}$ was applied to ensure that these baselines would not be affected by the Sun as an extended source. Under this condition, we selected the following baselines: C23-C28, C23-C29, C23-C30, C23-C31, C23-C32, C24-C29, C24-C30, C24-31, C24-C32, C25-C30, C25-C31, C25-C32, C26-C31, C26-C32, C27-C32. The length of these baselines range from 1.88m to 3.38m.

\subsubsection{Phase Calibration}
The visibility data used in this work are calibrated (for details, see \citealt{zuocal}). Calibration is typically performed using sources Cygnus A (Cyg A), Cassiopeia A (Cas A), and Taurus A (Tau A). The calibrated visibility is ideally given by
\begin{equation}
    \label{eq2:calvis}
    V_{ij} = \frac{g_ig_j^*}{\tau}\int \dif^2 \vec{n} \dif t\ \ |A(\vec{n})|^2 e^{-\frac{2\pi i}{\lambda} (\vec{r}_j - \vec{r}_i)\cdot \vec{n}}I(\vec{n}) + <\eta_i \eta_j^*>
\end{equation}
where $g_i$ is the complex gain of receiver feed $i$, $\tau$ is the integration time, $I(\vec{n})$ is the sky intensity distribution, and $\vec{r}_i$ is the position vector of feed $i$, $\eta_i$ is the random noise of the receiver $i$, the average in the last noise term is assumed to vanish except for $i=j$.
Since the measured gain amplitude can be normalized to eliminate the effect in measuring primary beam, only phase calibration is performed here. For the selected baseline, the expected visibility phase is calculated by treating the Sun as a point source when it reaches transit. Instead of calibrating the gain $g_i$ for each individual feed, we calibrated the gain of the baseline $(i,j)$. This calculated phase is then compared with the actual visibility to derive the required calibration phase. 

\subsubsection{Beamforming}
The Sun can be treated as a point source for the selected short baselines. Since the solar radio flux is very strong, we can reasonably ignore the influence of other sources, and \eref{eq2:calvis} is reduced to the form
\begin{equation}
    \label{eq2:pointvis}
    V_{ij}^{\rm cal}(\vec{n}) = I_{\rm Sun}A^2(\vec{n}) e^{- \frac{2\pi i}{\lambda}\vec{n}\cdot(\vec{r}_j - \vec{r}_i)}
\end{equation}
where $\vec{r}_i$ is position vector of feed $i$, $I_{\rm Sun}$ is the flux of the Sun, and $\vec{n}$ is the unit vector pointing towards the sun. We can make a beamformed output from the visibilities as 
\begin{equation}
    S = \frac{1}{N}\sum_{(i,j) \in \mathcal{S}} V_{i,j}^{\rm cal} ~ e^{\frac{2\pi i}{\lambda}\vec{n}\cdot(\vec{r}_j - \vec{r}_i)}
\end{equation}
where $\mathcal{S}$ is the set of short baselines used for this purpose and $N$ is the number of selected baselines. The solar position $\vec{n}$ as a function of time is computed using the Pysolar Python library
\footnote{https://github.com/pingswept/pysolar}.
We averaged the data of selected baselines and adjacent days. This averaged beamformed visibility is a measurement of the average primary beam, which is different with primary beam of each individual feed.

\subsection{North-South Beam Measurement}
\begin{figure}[htbp]
    \centering
    \includegraphics[width=0.8\textwidth]{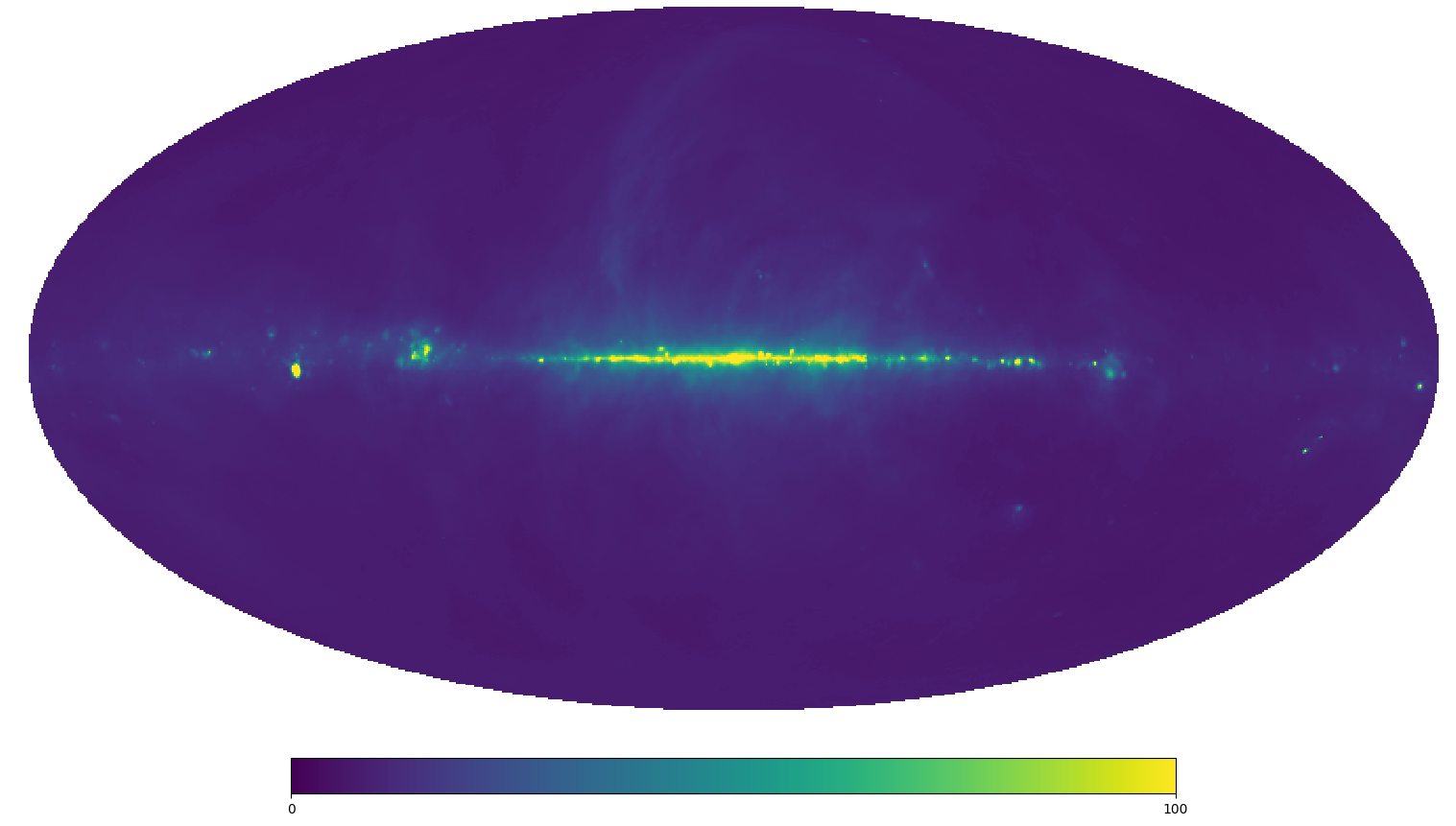} 
    \caption{The sky model generated using PyGDSM at the TCPA center frequency (750 MHz) in Galactic Coordinates .}
    \label{fig:skygdsm}
\end{figure}

To fit the primary beam in the North-South direction, we constructed a sky model and simulated the received visibilities. The primary beam in the North-South direction was treated as a fitting parameter.
Here we utilized a sky map developed by \cite{gdsm16}, which is displayed in \fref{fig:skygdsm} at the TCPA center frequency (750 MHz). However, since this sky map model is missing some point sources, we manually added a number of point sources, such as Cyg A and Cas A. 

The flux of Cas A is modeled as: 
\begin{equation} 
    \label{eq2:casflux} 
    S_{\text{Cas A}} = S_{0}(1-p)^n \left(\frac{\nu}{\nu_0}\right)^{\alpha} 
\end{equation} 
where $S_{\text{Cas A}}$ is the flux density and $\nu$ represents the observing frequency \cite{fluxform}. It changes over time \citep{72fluxdecrease}, so we utilized the datasets from \cite{cas19data} and \cite{cas15data} to fit the parameters in \eref{eq2:casflux}, thereby determining the flux of Cas A at the specific time of observation.

Our sky map model takes the following form
\begin{equation}
    \label{eq2:skymodule}
    I_{\rm sky}(\vec{n}) = I_{\rm GDSM2016}(\vec{n}) + I_{\rm Cyg}\delta(\vec{n} - \vec{n}_{\rm Cyg}) + I_{\rm Cas}\delta(\vec{n} - \vec{n}_{\rm Cas})
\end{equation}
where $\delta$ denotes the Dirac function. 
Based on \eref{eq2:beamNS}, we can express the visibility as
\begin{equation}
    \label{eq2:simulation}
    V_{ij}^{\rm simu} = \beta\int \dif^2 \vec{n}\ \ A^2_{NS}(\vec{n})A^{2}_{EW}(\vec{n}) I_{\rm sky}(\vec{n})
\end{equation}
We can assume that the difference between $V_{ij}^{\rm simu}$ and $V_{ij}$ is thermal noise and follows a Gaussian distribution. The parameter $\beta$ appears in \eref{eq2:simulation} because the sky model might differ in flux by a factor due to issues with flux and calibration. If the sky model matches the actual situation, its value should be 1. For computational convenience here, we transform the form of $A_{\rm NS}$ to
\begin{equation}
    \label{eq2:beamalpha}
    A_{\rm NS}(\theta) = e^{-\frac{\theta^2}{\theta_{\rm NS}^2}} = (A_{\rm NS}^{\rm simu})^{\frac{1+\alpha}{2}}
\end{equation}
Here, $A_{\rm NS}^{\rm simu}$ is the $A_{\rm NS}$ corresponding to the best-fit $\theta_{\rm NS}$ obtained from electromagnetic simulations.


For extended sources such as the Galaxy, the signal diminishes as the baseline length  ($L$) increases. For long baselines, the strength of the bright point sources and the extended sources would differ too much, so we select a group of short baselines. For the TCPA configuration, all short baselines are those connecting receiver feeds on the same cylinder and oriented in the north-south direction. Finally, We have selected 12 similar short baselines located on the same cylinder. Under this condition, we selected the following baselines A1-A9, A2-A10, A3-A11, A4-A12, A5-A13, A6-A14, A7-A15, A8-A16, A9-A17, A10-A18, A11-A19, A12-A20. The length of these baselines are all 3.2m.

We adopt a multivariate Gaussian likelihood function of the form
\begin{equation}
    \label{eq2:likelihood}
    \log(L(\alpha, \beta)) \sim (V_{ij} - V_{ij}^{\rm simu}) (V_{ij} - V_{ij}^{\rm simu})^* / \sigma^2
\end{equation}
Here, we assume that $\alpha$ and $\beta$ satisfy a uniform distribution, and $\sigma$ is the standard deviation of the thermal noise, which is determined by observing an `empty' sky region devoid of bright radio source.

\section{Results}
\label{section:results}
We now apply the methods described above to the observational data of the TCPA. The data sets we use are listed in Table \ref{tab:testdata}.

\subsection{Results of Phase Calibration}

\begin{figure}[ht!]
    \centering
    \includegraphics[width=0.9\textwidth]{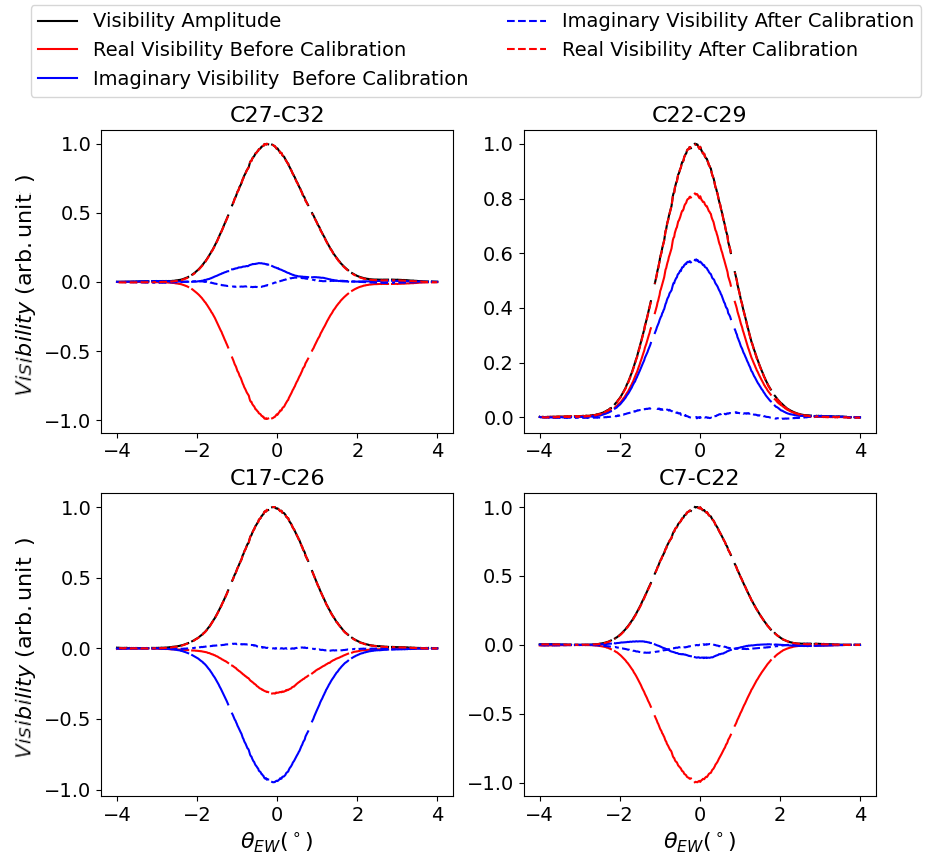} 
    \caption{Comparison of the visibility for four baselines before (solid curve) and after (dashed curve) phase calibration using the Sun. Red, blue and black curves correspond to the real, imaginary parts and amplitude.  {The discontinuities in the segments are the noise source signal for calibration which have been masked.}
    }
    \label{fig:phasecal}
\end{figure}

We show the effect of phase calibration on the beamforming result for one baseline in \fref{fig:phasecal}. Theoretically, if the phase calibration is perfect, the beamformed output should be a real number, with zero imaginary part. However, after direct beamforming, we observe a non-zero imaginary part that varies synchronously with the magnitude, though it is smaller than the real part. This may be caused by a change in the instrument phase of the gain since the time of last calibration, due to the change in temperature.  

To mitigate this effect, we can also make an instrument phase calibration during the transit of the Sun, by treating it as a point source, and assuming the instrument phase to be a constant during the transit. After phase calibration, we can see that the imaginary part has been reduced significantly.  This residual imaginary part may be due to error in the calibration, e.g. there could be variation of the instrument phase during the solar transit time itself, when the reflected sunlight heats up the feed, or due to the fact that the Sun is not exactly a point source. Fortunately, as we can see the amplitude is not much affected by this phase error.

\subsection{Test of the Separable Form of Primary Beam Model}

Next, we test our assumption that the primary beam is of the separable form, by observing the transit of the Sun at different elevation angles, which is shown in \tref{tab:testdata}. Since the primary beam of Tianlai is very narrow, the beamforming results within a single day can be considered to represent the east-west primary beam. 

However, since the Sun is an extended source, the measured data is theoretically a convolution of the E-W beam $A_\mathrm{EW}$ and the solar brightness distribution. We evaluated this effect by convolving our beam model with a uniform solar disk and performing a deconvolution on the observational data. Our analysis shows that this convolution effect is negligible compared to our measurement precision. 

{ Furthermore, in the actual measurements, the center of the measured E-W beam may exhibit a small offset due to the error in feed orientation alignment \citep{LiCylDes}. Here we show the result after constraining the center of the measured E-W beam to the pointing axis.}

\begin{table}[ht] 
    \centering 
    \caption{Summary of the datasets used in this work. {The parameter $\theta_z$ is defined as the minimum angle between the Sun and the local zenith.}} 
    \label{tab:testdata} 

    \begin{tabular}{|c|c|c|c|} 
        \hline 
        Data set ID & Start date & End date & {Solar Zenith Angle $\theta_z$}  \\ 
        \hline 
        1 & 2016-09-27 & 2016-10-02 & 45.8$^\circ$ \\ 
        \hline
        2 & 2018-03-22 & 2018-03-31 & 43.4$^\circ$ \\ 
        \hline
        3 & 2023-06-21 & 2023-06-26 & 20.6$^\circ$ \\ 
        \hline 
        4 & 2023-09-06 & 2023-09-10 & 37.7$^\circ$ \\ 
        \hline
        5 & 2024-11-27 & 2024-12-09 & 62.5$^\circ$ \\ 
        \hline
        6 & 2018-01-13 & 2018-01-16 & 65.5$^\circ$ \\ 
        \hline
        7 & 2018-03-31 & 2018-04-02 & 39.9$^\circ$ \\ 
        \hline
    \end{tabular}
\end{table}

\begin{figure}[htbp]
    \centering
    \includegraphics[width=0.6\textwidth]{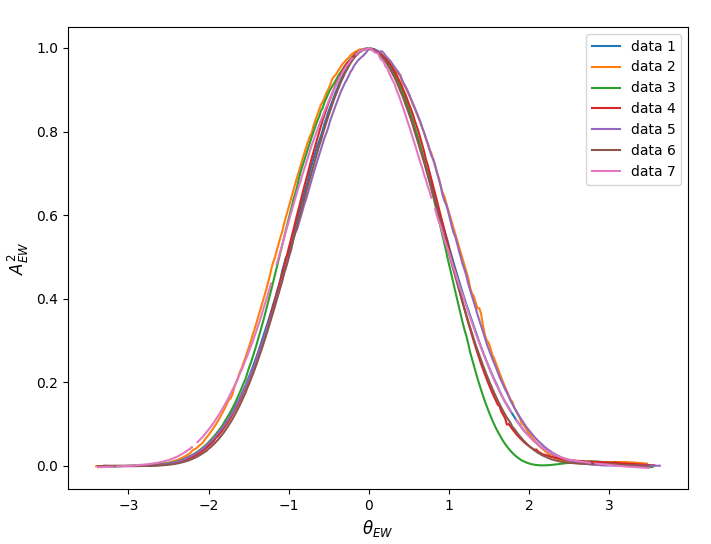} 
    \caption{E-W primary beam measurement results derived from different datasets for the center frequency.}
    \label{fig:testbeam}
\end{figure}

\begin{figure}[htbp]
    \centering
    \includegraphics[width=0.6\textwidth]{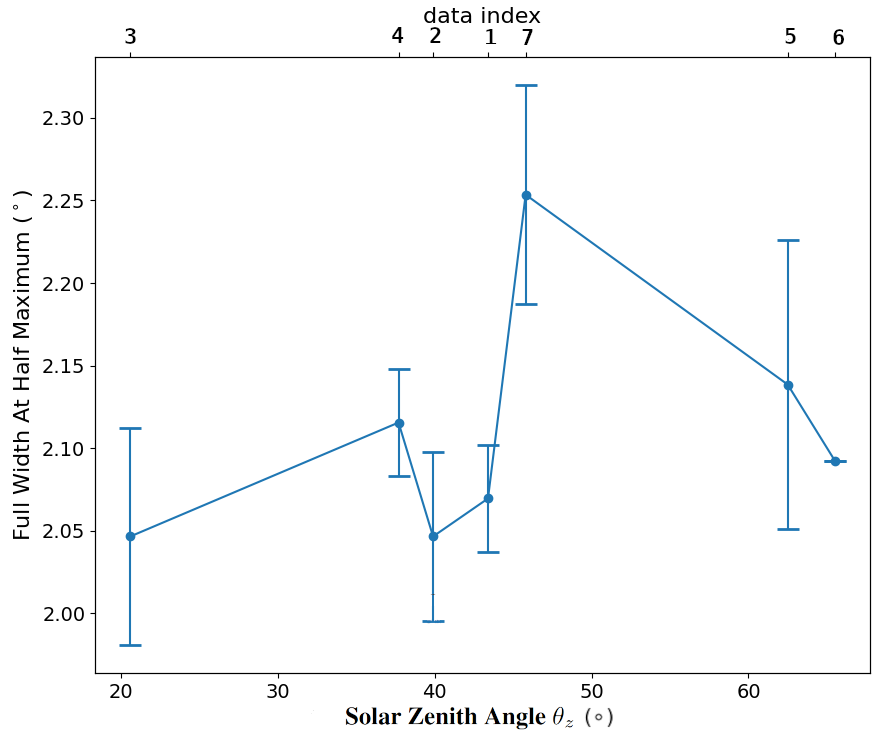} 
    \caption{The Calculated FWHM values with associated 3$\sigma$ uncertainties. The horizontal axis follows the solar transit angle, ordered chronologically to illustrate the beam's stability during the transit.}
    \label{fig:testFWHM}
\end{figure}

\begin{figure}[htbp]
    \centering
    \includegraphics[width=0.9\textwidth, trim=0cm 0cm 0cm 1.5cm, clip]{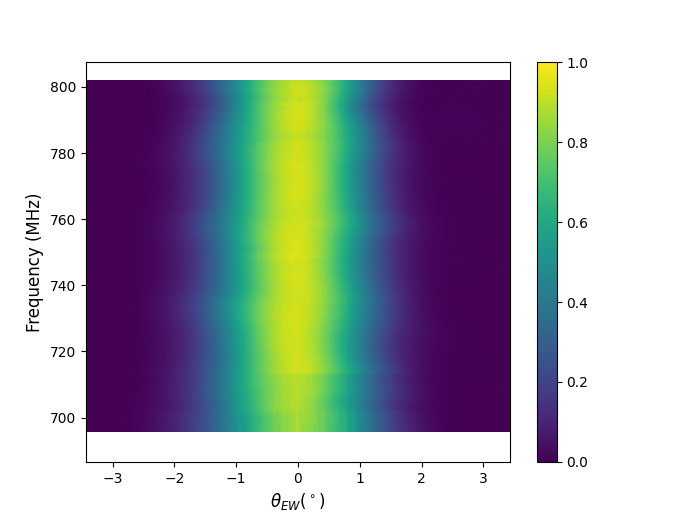} 
    \caption{The East-West primary beam profiles across different frequencies.}
    \label{fig:testfreq}
\end{figure}

The East-West primary beam profile measured with the different datasets which correspond to different elevation angles is shown in \fref{fig:testbeam}. The Full Width at Half Maximum (FWHM) is calculated for each transit and shown in \fref{fig:testFWHM}. It is evident that the beam width remains largely invariant regardless of the Sun's position. These profiles are very similar, with nearly identical width, as predicted by the separable form we assumed for the beam profile. To reduce random errors, the E–W primary beams obtained from different data sets were averaged to derive the final beam profile. In \fref{fig:testfreq}, we show the measured beam profile at different frequencies.

\subsection{Measurement of the North-South Primary Beam}

With the East-West parameters derived from the above measurements, we now fit the North-South factor of the primary beam. The model we adopt has been described in \sref{section:methods}. 


We fit the beam by using the visibility data from a group of relatively short North-South baselines including A1-A9, A2-A10, A4-A12. Because for extended sources such as the Galaxy, the signal diminishes as the baseline length  ($L$) increases. For long baselines, the strength of the bright point sources and the extended sources would differ too much. For the TCPA configuration, all short baselines are those connecting receiver feeds on the same cylinder and oriented in the north-south direction. We have selected 12 similar short baselines located on the A cylindrical surface.
We fit the primary beam parameters at the different frequencies separately. 

\begin{figure}[htbp]
    \centering
    \includegraphics[width=0.8\textwidth]{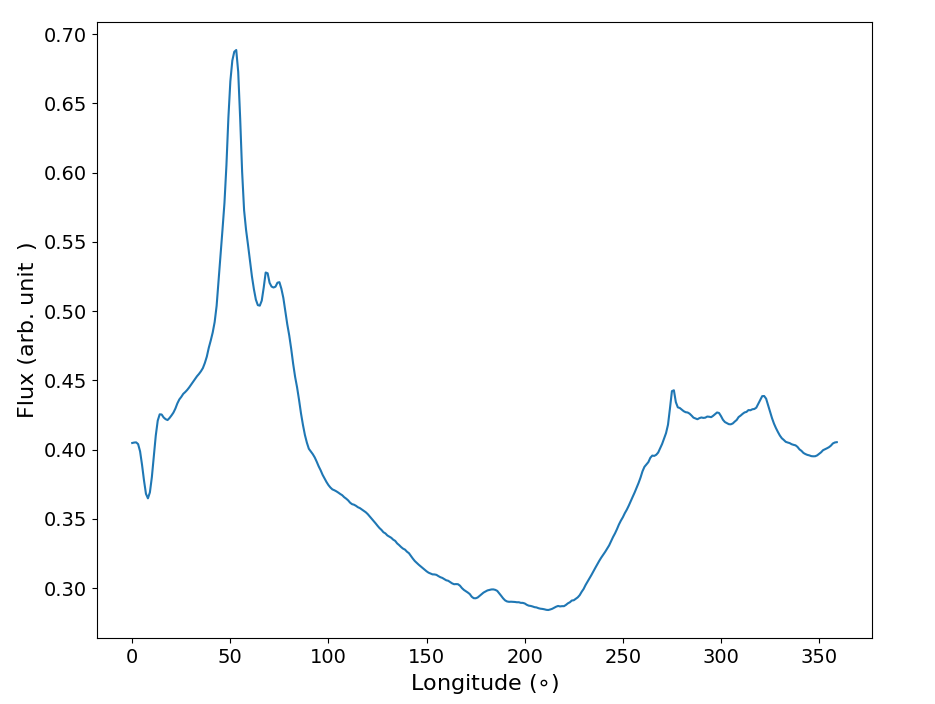} 
    \includegraphics[width=0.8\textwidth]{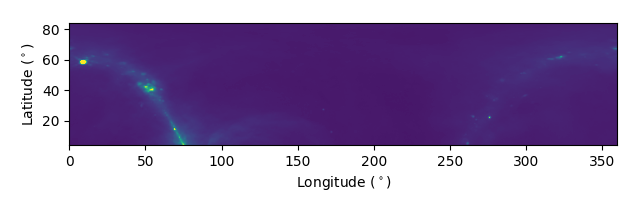} 
    \caption{the sky map model and simulated flux from the TCPA perspective when $ | \theta_{\rm NS} |<40^\circ$}
    \label{fig:tianlai_lat}
\end{figure}

The sources that have a constraining effect in our data are primarily Cyg A, Cas A, and the Galactic plane. Other sources are not sufficiently bright to be separated from the noise. As noted in the previous section, we use the PyGDSM sky map to model the intensity distribution of the Galactic plane, as shown in \fref{fig:tianlai_lat}. The error in this model would cause an error in the calibration of our beam profile. 


\begin{figure}[htbp]
    \centering
    \includegraphics[width=0.8\textwidth]{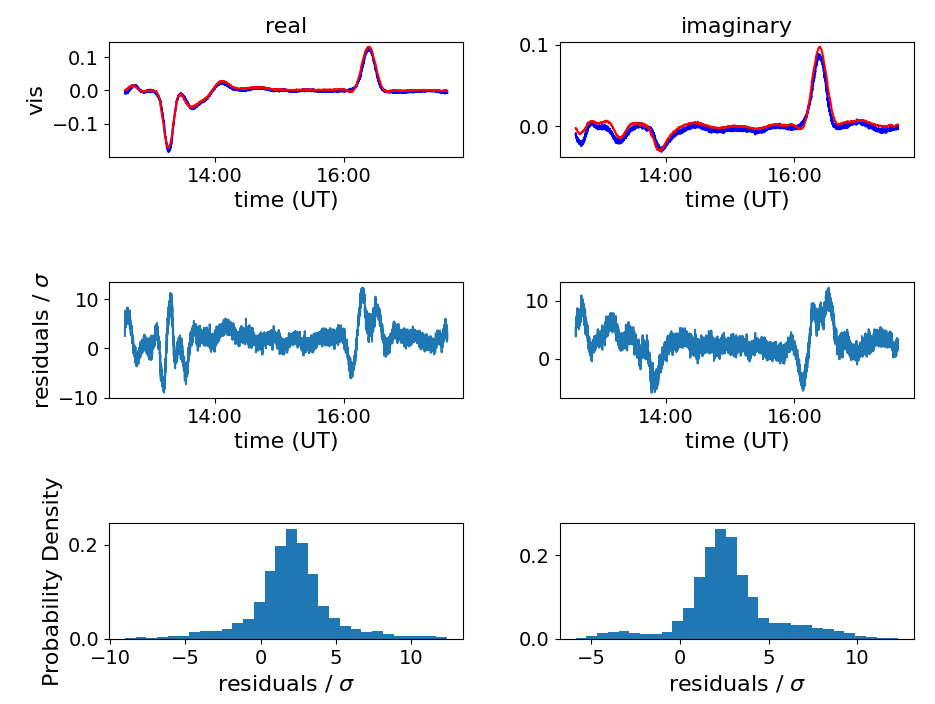} 
    \caption{Top: The observed data in data set 1 (blue line) and the simulated signal using the best-fit parameters (red line) for the selected short baseline; Middle: residuals; Bottom: Distribution of residuals 
    }
    \label{fig:data_fit}
\end{figure}

In \fref{fig:data_fit} we show the visibility for a short north-south baseline at 750 MHz. We also plot the simulated result for the best-fit parameters $\alpha = 0.116 \pm 0.002, \beta = 1.45 \pm 0.002$. We see the fit is pretty good. 



\begin{figure}[htbp]
    \centering
    \includegraphics[width=0.8\textwidth,trim=0 0 0 0cm, clip]{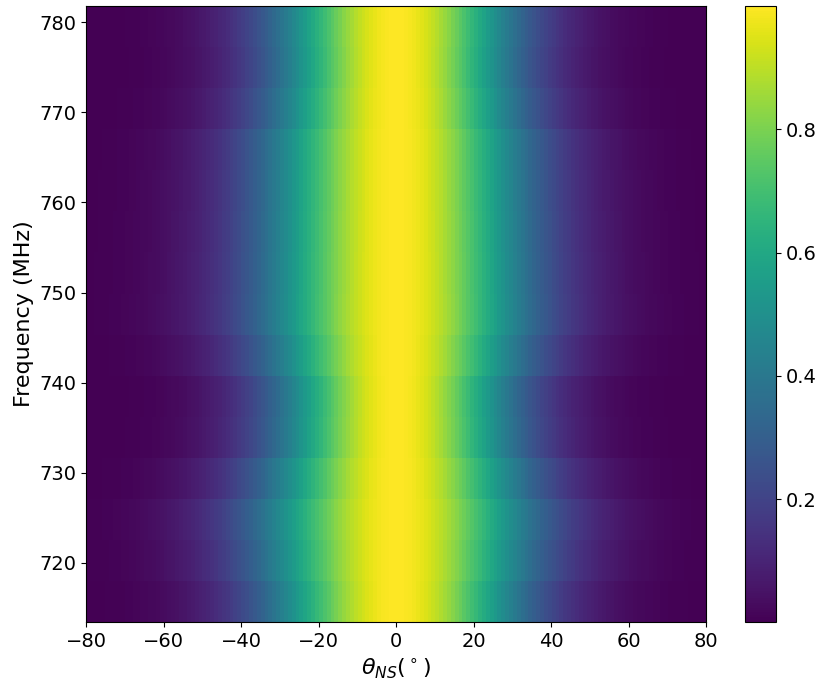} 
    \caption{The morphology of the primary beam along the North-South direction across different frequencies. }
    \label{fig:NS-beam}
\end{figure}


The shapes of the north-south primary beams for different frequencies are shown in \fref{fig:NS-beam}.

\subsection{Comparison {with} Drone-based Measurements}
\label{comparison}
\begin{figure}[htbp]
    \centering
    \includegraphics[width=1.1\textwidth]{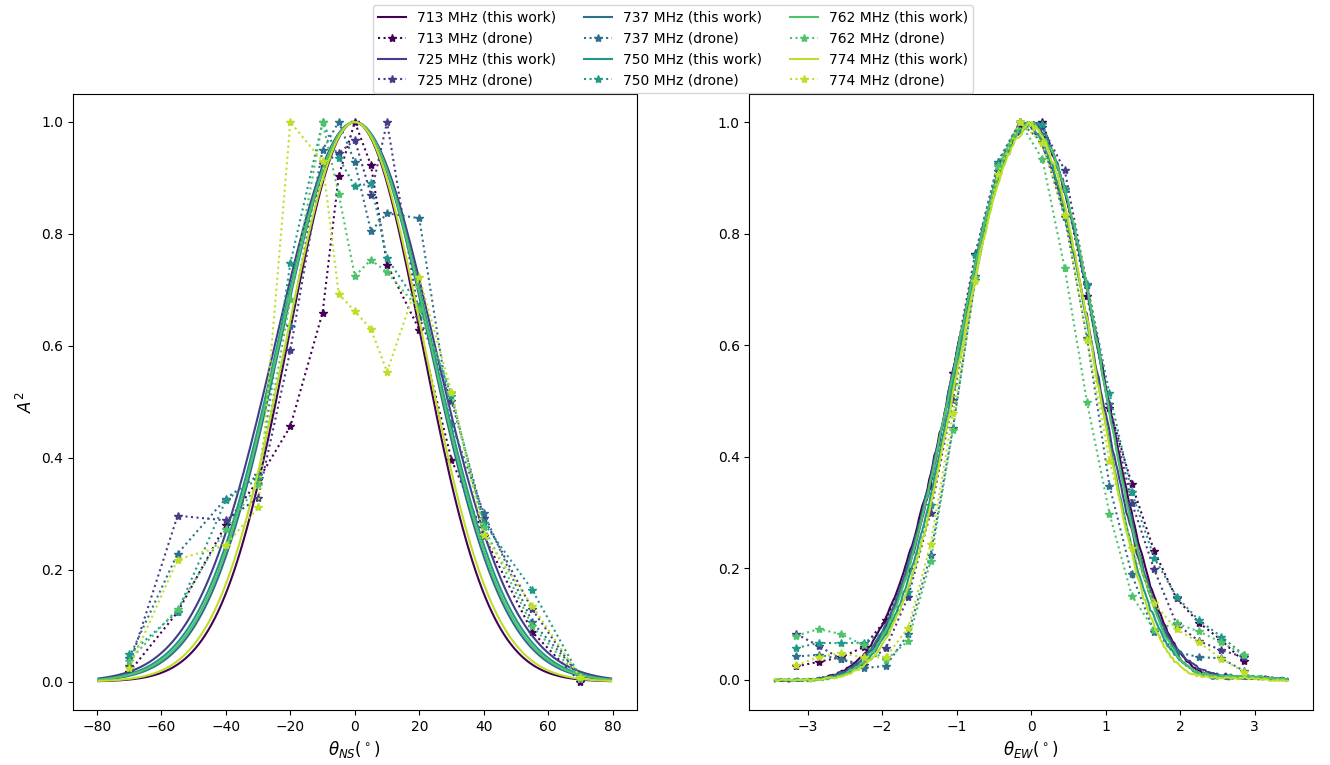} 
    \caption{Comparison of the results at the center frequency. Left: Comparison of the primary beam in the North-South (N-S) direction; Right: Comparison of the primary beam in the East-West (E-W) direction. The baselines used here are consistent with the method described in \sref{section:methods}.
    }
    \label{fig:comparison}
\end{figure}

Here, we compare the model beam profile we obtained with the drone-based measurements \citep{LiDrone}. The comparison plots are shown in \fref{fig:comparison} for a number of different frequencies. 

As can be seen in the plot, for the profile along the East-West direction (right panel), the beam measured by the drone is very close to a Gaussian, and it agrees relatively well with our measurements.

Along the North-South direction (left panel), the drone measured beam can deviate significantly from the Gaussian shape. Such discrepancies may arise from errors inherent in the drone-based methodology, such as tilting due to wind influence or inaccuracies in the drone's own beam profile. Furthermore, ground reflections from nearby hills may also interfere with the measurement results \citep{LiDrone}. Alternatively, it is possible that the actual beam shape intrinsically deviates from our model, which only captures a general profile. However, the width of these beam profiles does approximate the ones we determine in this work. 



\begin{figure}[htbp]
    \centering
    \includegraphics[width=1\textwidth]{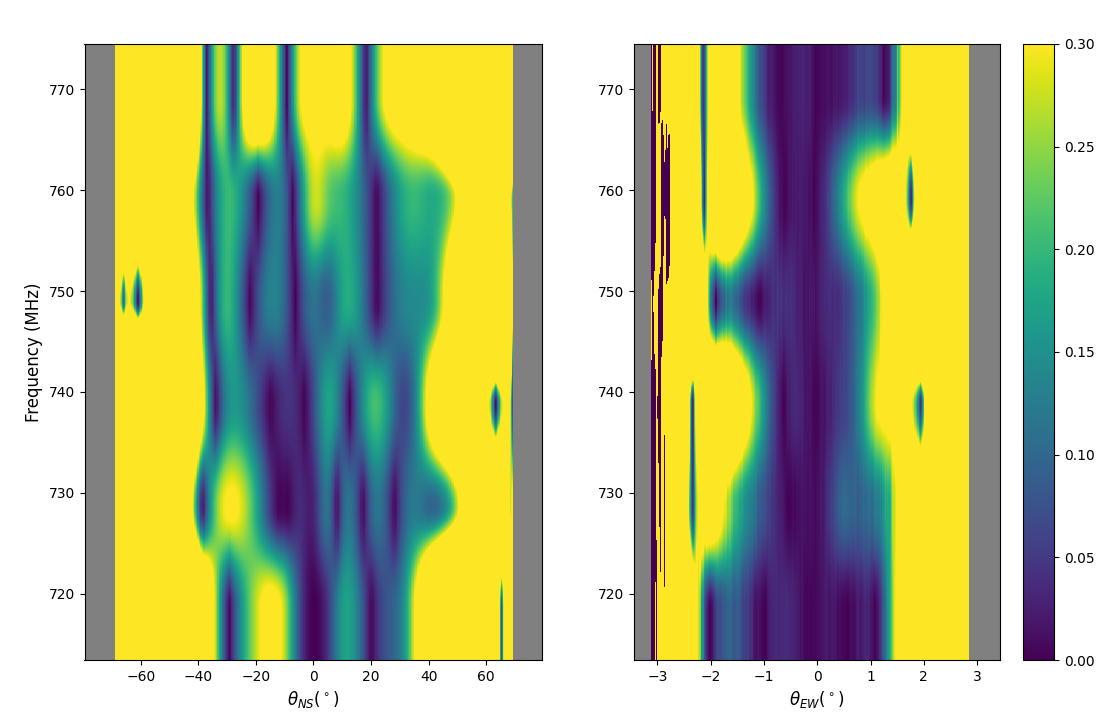} 
    \caption{Relative differences between the drone-based measurements and our model results, $\frac{ | A^2_{\text{drone}} - A^2_{\text{this work}} | }{A^2_{\text{this work}}}$ . The left panel shows the N-S primary beam, while the right panel displays the E-W primary beam.}
    \label{fig:comparison_error}
\end{figure}

The relative error between the two methods, is shown  in \fref{fig:comparison_error} across the observed frequency range. This comparison shows the difference between the drone measurement and our data-fitted model. 
The general behavior is similar, though the exact deviation varies over the frequency. The relative differences are smaller near the center of the beams, though they become large at large angles, where the beam profiles fall to small values and measurement errors also become large. Fortunately, at these large angles the contribution to the data is also smaller.

\section{Discussions}
\label{section:discussion}

In this work, our measurement results are related to our model. However, the actual primary beam may deviate from our model predictions. For example, we expect north-south symmetry in the east-west primary beam within the model presented in \sref{section:methods}, yet the actual beam profile may deviate from these theoretical assumptions. In this section, we discuss the systematic errors inherent in our methodology.

The errors in our methodology may arise from several sources: 1) Modeling error in the instruments, e.g. the form of the actual beam profile is not a Gaussian function, its peak is off-center and its profile is not even strictly symmetric. 2) Modeling error of the sky. We have used the model of Cyg A, Cas A and the Galactic plane for this calibration; errors in their flux will affect our results. 3) Measurement errors, e.g. those arising from the noise and fluctuations during the measurement process. Note that if we compare our results to other measurements, they will also have their own sources of errors. For example, in the case of beam measurement with the drone, while they do not use the sky model, there are still errors due to the model of instruments and noise. 


\begin{figure}[htbp]
    \centering
    \includegraphics[width=0.8\textwidth]{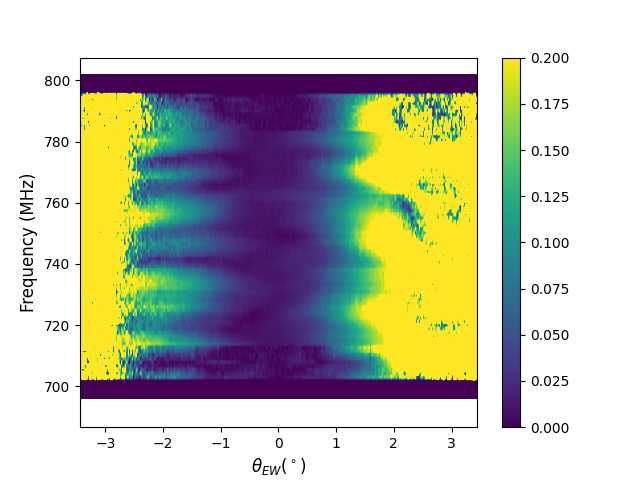} 
    \caption{Relative error estimation of the E-W primary beam derived from resampling.}
    \label{fig:EW_beam_error}
\end{figure}

To estimate the model errors introduced by our approach in the E-W profile, we resampled the primary beam in the E-W direction to obtain the relative error of the measurements. This approach does not require the data errors to follow any specific distribution, making it model-independent. The results are shown in \fref{fig:EW_beam_error}. 
It can be observed that within the primary beam range, the relative error of the 
measurement is within 10\%, suggesting that the uncertainty of our results is relatively small.

For the error in the North-South profile, we have already compared our measurement with the drone, but bear in mind that the drone measurement also has its own error. Here we assess the impact of the sky model, in particular the error due to the flux model of the Cyg A and Cas A. The relative error of the observed flux is within or around 5\%, which we expect will not significantly affect our results. Since the flux densities of Cyg A and Cas A far exceed those of other celestial sources, the measurement errors induced by their flux are expected to be the most significant, followed by the uncertainty in the Galactic plane emission. These two components constitute the primary contributors to our sky model error. Although random measurement errors may also affect the results, we assume they are uncorrelated, and by averaging over multiple baselines, this noise can be suppressed to a negligible level.



\section{Conclusion}
\label{section:conclusions}
In this work, we characterized the primary beam pattern of the Tianlai Cylinder array by using a combination of the Sun as a calibration source and a sky map model. Under the assumption that the primary beam is of a separable form, we directly measured the East-West (E-W) beam profile using the Sun as a point source, while simultaneously deriving the best-fit North-South (N-S) beam profile through fitting. The validity of the separability assumption was subsequently verified by comparing data sets across different time intervals. Furthermore, we conducted a comparative analysis between our current results and the previous findings reported in \cite{LiDrone} and \cite{Sun_2022}, showing that the results are largely consistent. Finally, to evaluate the systematic uncertainties of our methodology, we employed a resampling technique to analyze the procedural errors.

\begin{acknowledgements}
We thank the referee for comments and suggestions that improved the paper. This work is supported by the National SKA Program of China (Nos. 2022SKA0110100 and 2022SKA0110101), the National Natural Science Foundation of China (NSFC) International (Regional) Cooperation and Exchange Project (No. 12361141814), the NSFC (No.12303004,12203061,12473094,12273070.),the NSFC innovation group grant 12421003, the Specialized Research Fund for State Key Laboratory of Radio Astronomy and Technology, and the National Astronomical Observatories, Chinese Academy of Science (No. E5ZB0901).This work is also supported by science research grants from the China Manned Space Project with grant Nos. CMS-CSST-2021-B01 and CMS- CSST-2021-A01.
\end{acknowledgements}

\bibliographystyle{raa}
\bibliography{2026-0218}



\end{document}